# ACCDB: A Collection of Chemistry DataBases for Broad Computational Purposes

**Pierpaolo Morgante,[1] Roberto Peverati[1*]**

[1] P. Morgante, R. Peverati
*Chemistry Program, Florida Institute of Technology, 150 W. University Blvd., Melbourne, FL 32901, United States.*
*E-mail: rpeverati@fit.edu*

**ABSTRACT.** The importance of databases of reliable and accurate data in chemistry has substantially increased in the past two decades. Their main usage is to parametrize electronic structure theory methods, and to assess their capabilities and accuracy for a broad set of chemical problems. The collection we present here—ACCDB—includes data from 16 different research groups, for a total of 44,931 unique reference data points, all at a level of theory significantly higher than DFT, and covering most of the periodic table. It is composed of five databases taken from literature (GMTKN, MGCDB84, Minnesota2015, DP284, and W4-17), two newly developed reaction energy databases (W4-17-RE, and MN-RE), and a new collection of databases containing transition metals. A set of expandable software tools for its manipulation is also presented here for the first time, as well as a case study where ACCDB is used for benchmarking commercial CPUs for chemistry calculations.

## Introduction

Databases of atomic and molecular energies have always been an important tool in computational chemistry for the assessment and parametrization of semi-empirical methods.[1,2] Large sets of experimental data of small molecules became increasingly important with the development of methods for the calculation of heats of formation of molecules, starting from the early days of semi-empirical methods (for a historical perspective see Ref. [3]), to the Gaussian composite methods of Curtiss and co-workers,[4] and the multi-coefficient correlation methods of Truhlar and co-workers.[5] More recently, the development of high-accuracy *ab initio* calculation methods has shifted the interest from databases of experimental data to databases of purely calculated data. The advantages of the latter for the assessment and parametrization of semi-empirical electronic structure methods is clear, since they include data that are directly comparable between calculations, without the need of corrections for experimental conditions, such as zero-point energies (in most cases), thermal corrections (for heats of formation), anharmonicity effects (for vibrational frequencies), vector-relativistic effects (spin-orbit couplings), *etc*. Moreover, high-level calculated data do not suffer from experimental errors. Albeit they do still suffer from the computational accuracy error of the level of theory that is used to obtain them, such error is easier to assess and to control than the stochastic experimental error, and in many cases, it is preferable.

The importance of computational chemistry databases containing high-level data has grown exponentially in the last two decades because of the advent of semi-empirical exchange-correlation (xc) functionals in density functional theory (DFT). While the necessity of extremely large databases for parametrization of new xc functionals—and the increasing number of parameters in the functional forms—is a controversial topic that is outside the scope of this work (see for example refs. [6,7], and for a recent debate, see [8-10]), we believe that databases will undoubtedly remain a fundamental tool for the assessment of the applicability, accuracy, and reliability of new and existing xc functionals in the years to come, regardless of the underlying wars between optimization philosophies (from first principle vs. parametrized). It is objectively true that more parameters in the xc functionals require more data in the training set (a statistical necessity to avoid over-training), but the availability of larger training sets does not necessarily translate into more parametrized functionals, nor into more parameters in a functional.[11]

The modern use of computational chemistry databases containing a large number of high-level calculated data is not limited to the parametrization and assessment of semi-empirical composite or DFT methods: other interesting applications include validation of high-accuracy *ab initio* method,[12] benchmarking of software or hardware,[13-15] and the application of modern data mining techniques to chemistry, such as artificial intelligence and machine learning.[16]

In light of this increasing importance of large sets of calculated data with high accuracy in chemistry, we present here a collection of almost 45k unique reference data points (with ~ 37k of them presented here for the first time), all at a level of theory significantly higher than DFT. This collection—that we named ACCDB—includes data from 16 different research groups, a total of more than 10k atomic and molecular structures files, as well as a set of software tools for their manipulation, automation of corresponding jobs, and statistical analysis. Its primary difference with respect to other recent large databases of chemical compounds[17-20] is the high level of accuracy at which our data are obtained. Another key aspect of ACCDB is its broad applicability to different areas of chemistry, including—but not limited to—both main-group and transition-metals thermochemistry, non-covalent interactions, and chemical kinetics.

In the next two section we describe the structure of ACCDB, including details for all the databases that compose it, and for the automation tools that we developed to simplify the management of such a large quantity of data. Before the conclusion, we also present a case study where ACCDB is used to benchmark the performance of three commercial CPUs recently introduced onto the market (as of August 2018) by AMD, for routine computational chemistry calculation.

## Structure of the database

ACCDB includes five different databases from five different sources: MGCDB84,[21] GMTKN,[22,23] Minnesota,[24,25] DP284,[26,27] and W4-17.[12] The data from these sources cover primarily main group elements. Some transition metals (TMs) are present in the Minnesota database, but we significantly expand this number by introducing here a database of reactions involving first-, second-, and third-row TMs, as well as elements from the second transition. In addition to these "traditional" data points, we also introduce here two new databases obtained using automatic generation of reaction energies,[28] containing further 36,275 "non-traditional" reference data points.

In total, ACCDB includes 191 subsets and 44,931 data points. A brief description of each database is presented below, and a summary is in Table 1. We suggest the user of each pre-existing database to refer to the original publications for more information on the subsets and to give proper credit to its primary authors.

| Table 1. Summary of all databases included in ACCDB.[a] | | | | |
|---|---|---|---|---|
| Name of the database: | Brief description of what is included in the database: | Number of Structures: | Unique Reference Data Points: | Ref. |
| MGCDB84 | Main Group Chemistry DataBase | 5,931 | 4,985 | 21 |
| GMTKN | General Main Group, Thermochemistry, Kinetics, and Non-covalent interactions and Mindless Benchmarks | 2,639 | 1,664 | 22,23 |
| Minnesota | Thermochemistry, Kinetics, Non-Covalent interactions (Database2015, Database2015A, and Database2015B) | 719 | 471 | 24,25,29 |
| *- MN-RE* | *Automatically Generated Reaction Energies from Minnesota 2015B* | - | *9,135* | *This work* |
| DP284 | Dipole Moments and Polarizabilities | 181 | 284 | 26,27 |
| W4-17 | Total Atomization Energies | 215 | 1,042 | 12 |
| *- W4-17-RE* | *Automatically Generated Reaction Energies from W4-17* | - | *27,140* | *This work* |
| Metals&EE | Collection of subsetscontaining metals and excitation energies | 364 | 210 | 30-40 |
| **ACCDB** | | 10,049 | 8,656 (44,931)[b] | |

[a] For further details of all subsets of each database—and clarifications on the corresponding reference—see the supporting information.
[b] The number in parentheses includes the "non-traditional" RE databases.

***MGCDB84.*** The Main Group Chemistry DataBase has been introduced by Mardirossian and Head-Gordon.[21] It is the largest database included in ACCDB, with 84 subsets, 5,931 single-point geometries, and 4,985 reference data. Part of it was used as training set for new xc functionals,[11,41-43] while it has been used as benchmark set in its entirety, for assessing the performance of the Minnesota functionals,[44] as well as about 200 more functionals.[21] Its main focus areas are: non-covalent interactions, which make up almost half of the database (2,647 data), thermochemistry (1,205 data), isomerization energies (910 data) and kinetics (barrier heights: 206 data). Additionally, the electronic energies of the first 18 atoms are also included. Quite recently, a statistically significant version of the MGCDB84 database, called MG8, has been proposed by Bun Chan.[45]

***GMTKN.*** This database is composed by the GMTKN55 database,[22] and the MB08-165 database,[23] both from Goerigk's and Grimme's groups. GMTKN55 is the extension of two previously published databases by Goerigk *et al.*, called GMTKN24[46] and GMTKN30.[47,48] It includes 55 subsets, 2,459 single-point geometries, and 1,499 relative energies. The four areas of interest are: non-covalent interactions (both inter- and intra-molecular interactions, 595 data), basic properties (atomization energies, electron and proton affinities, dissociation energies of various compounds, 467 data), thermochemistry (reaction energies and isomerization reactions, 243 data), and kinetics (barrier heights, 194 data). MB08-165 is a database containing 165 "Mindless Benchmark", obtained with randomly-generated, artificial molecules each containing 8 atoms.[23] This database was included in both GMTKN24 and GMTKN30, but it was replaced in GMTKN55 by a new database of 43 artificial molecules of 16 atoms each (MB16-43). Because of its relevance as a chemically "unbiased" test set (the molecules are not real ones, reducing the possibility of forecasting the outcome of the calculations), we have decided to keep MB08-165 in our collection, in addition to GMTKN55. As a comprehensive example of the usage of GMTKN55 in the context of DFT, Goerigk *et al.*[22] performed a rigorous analysis of 217 xc functionals (all with and without semi-empirical dispersion corrections). Very recently, Goerigk and his group also used GMTKN55 to assess the performance of 29 double-hybrid functionals. [49] In a similar fashion to what was done by Bun Chan for MGCDB84, a "diet" version of the GMTKN55 database was recently proposed by Tim Gould.[50]

***Minnesota.*** The Minnesota database was developed in Truhlar's group for the parametrization of new xc functionals. We included in ACCDB the molecular data in Databases 2015,[29] Database2015A,[24] and Database2015B[25] (we do not include in ACCDB the geometry optimizations and solid-state sets). The first version of Database2015 is updated in both Database2015A—which includes 32 subsets, 652 geometries, and 422 reference energies—, and Database2015B—which includes 34 subsets, 719 geometries, and 471 reference energies. The focus areas are: bond energies (MGBE150 and TMBE33; 40% of the reference data), non-covalent interactions (NC87; 20% of the data), barrier heights (BH76; 16% of the data), thermochemistry (isomerization energies, excitation energies, hydrocarbon thermochemistry; 15% of the data), and basic properties (ionization potentials, atomic energies; 9% of the data).

***DP284.*** This database is a collection of two recent sets introduced by Hait and Head-Gordon, and it is comprised of 181 structures of small molecules. It includes 152 reference values for dipole moments,[26] and 132 reference data for polarizabilities,[27] obtained at the CCSD(T)/CBS level.

***W4-17.*** W4-17[12] is an extension of two previous databases—namely W4-08,[51] and W4-11[52]—developed by Martin's and Karton's groups. W4-17 includes 203 total atomization energies of 215 first- and second-row molecules and radicals with up to eight non-hydrogen atoms. We have also decided to keep the reaction energies generated in the original paper for the W4-11 version of the database,[52] thus including 99 data points from the BDE99 subset, 707 reference data for the HAT707 subset, 20 data points from ISOMERIZATION20, and 13 reference data from SN13. Each of these subsets have been generated using molecules as that are in W4-17, therefore no additional computation is needed to evaluate them. The reference energies for W4-17 and its subsets have been obtained using the highly accurate Weizmann-4 computational protocol, and they are guaranteed within a 3σ confidence intervals of 1 kJ mol$^{-1}$.

***W4-17-RE and MN-RE.*** These databases are presented here for the first time. Each of them includes reaction energies that are automatically generated from the W4-17 and Minnesota databases, using the *autoRE perl* script provided in Ref. [28] This program generates all the stoichiometrically feasible reactions of the form

$$A + B \rightarrow C + D,$$

from a corresponding list of atomization energy. The program automatically excludes all redundancies, double counting of reverse reactions, and trivial isomerizations. In this case, we started from the entire W4-17 database, and all atomization energies in the Minnesota Database2015B, to obtain the W4-17-RE, with 27,140 unique reference data, and the MN-RE, with 9,135 unique reference data. We refer to these databases as "non-traditional" because their usage for the parametrization and assessment of electronic structure methods is currently unexplored, and some statistical noise is expected, due to their large numbers of data. However, based on the work of Margraf *et al.*,[28] we expect low correlation between these sets and the corresponding atomization energies sets, supporting the inclusion of these "non-traditional" sets into ACCDB as independent databases.

***Metals and Excitation Energies.*** The Metals and Excitation Energies collection is also introduced here for the first time, with the goal to expand ACCDB to first-, second-, and third-row transition metals, as well as actinides from Th to Cm. Particular care should be given for this database when selecting an appropriate basis-set (in most cases, basis sets that include an effective core potentials are required), as well as for issues related with stability of the SCF solution, and proper treatment of spin-contamination. All of the data have been taken from different sources in the literature, and are divided into the following eight subsets:

*3d-SSIP30.* Spin-state (SS) energetics and ionization potentials (IP) of all 10 first-row 3d TMs (from Sc to Zn).[30] Spin states refers to the lowest-energy multiplicity-changing excitation energy for each species, and it includes data for both the neutral atom and the cations. Reference energies are experimental energies with spin-orbit coupling removed.

*4d-SSIP24.* This set is analogous to 3d-SSIP30 but for the first eight second-row 4d TMs (Y, Zr, Nb, Mo, Tc, Ru, Rh, Pd).[31]

*AIP28.* This database includes the ionization potentials of mono- and dioxides of actinides (Th to Cm).[32] It includes 42 geometries and 28 reference data. All data are calculated at the CASPT2/ANO-RCC (triple-zeta quality) level with a CAS(16,14) for the monoxides (14 orbitals for the metals, 2 for oxygen), and CAS(14,14) for the dioxide species. CASPT2 geometries are used as reference.

*TMBH23.* The Transition Metal Barrier Heights database includes reactions catalyzed by Zr, Mo, W and Re.[38-40] It includes 49 structures and 23 reaction energies: five are catalyzed by Zr, five are catalyzed by Mo, seven are catalyzed by W, and six are catalyzed by Re.

*LTMBH26.* The Late Transition Metals Barrier Heights database includes reactions catalyzed by Au, Pt, and Ir.[33] It includes 40 structures, and 26 reaction energies: two are catalyzed by Ir, two are catalyzed by Pt, and 22 are catalyzed by gold.

*MOR41.* The Metal Organic Reactions database of Grimme and co-workers[34] includes 41 data (95 structures and 13 different transition metals: Ti, Cr, Mn, Fe, Co, Ni, Mo, Ru, Rh, Pd, W, Ir, Pt, plus Al). All structures are carefully chosen to have single-reference character only.

*p-VR17.* This database includes valence (one electron goes from an *n*s to an *n*p orbital) and Rydberg (one electron goes from an *n*p to an (*n*+1)s orbital) excitations of different p-block elements and their mono-charged cations.[35] The elements included are: B, Al, Ga, F, Ne, Cl, Ar, Br and Kr; while the cations are: $B^+$, $C^+$, $Al^+$, $Si^+$, $Ga^+$, $Ge^+$, $Ne^+$, $Kr^+$, for a total of 34 atomic structures and 17 reference energies (9 valence + 8 Rydberg excitations).

*Por21.* This is a new database presented here for the first time. It includes spin states and binding energies data of porphyrin structures, which are ubiquitous in nature (the famous heme group is an example). It is divided into two subsets: PorSS11, which includes the spin-state energy differences of three Mn-porphyrins, one Co-porphyrin, and seven Fe-porphyrins, bonded to different ligands ($NH_3$, $OH^-$, $SH^-$);[36] and PorBE10, which includes the binding energies for the complexes between a model system of a heme group and three diatomic molecules: NO, CO, and $O_2$ (Figure 1).[37] Por21 includes 32 structures, and 21 reference energies obtained at the CASPT2 level (different active spaces have been used for each molecule in the database, their details are given in the original publications,[36,37] and are also reported in our online repository).

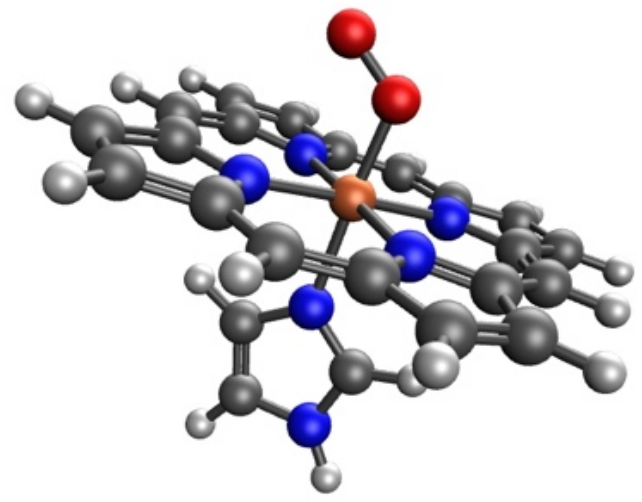

Figure 1. One of the structures in PorBE10, where a Fe-porphyrin is bonded to an imidazole moiety to mimic the binding environment in the heme group, and to an $O_2$ molecule. The binding energy is calculated with respect to the entire complex (including the imidazole moiety), and the $O_2$ molecule at infinite separation (carbon is black, hydrogen is white, nitrogen is blue, oxygen is red, iron is orange).

**Additional considerations.** For the sake of completeness, we have decided to collect all original databases and leave them untouched in ACCDB. For this reason, some of the subsets in some databases do overlap with each other, and the reference data for some data point in overlapping subsets are (on purpose) not always consistent. Hence, we advise care when using the entire collection, especially if the reduction of redundancies and/or of the number of calculations is a priority. For example, both GMTKN55 and MGCDB84 include the database W4-11, which is the previous version of W4-17 (with the latter having the most accurate reference data points). In addition, MGCDB84 heavily relies on data taken from GMTKN30—a preceding version of GMTKN55—and therefore the overlap between these two databases is substantial (and again, the most accurate data references have to be sought in the latter GMTKN55 database). Another complex situation is the DBH76 set of barrier heights, which is present in MGCDB84, as well as in GMTKN55, and also in Minnesota (albeit with different names, as HTBH38 and NHTBH38). As pointed out in the GMTKN55 paper,[22] the most recent—and most accurate—reference data for this subset are the one obtained by Goerigk and co-workers and included in GMTKN55. However, for consistency purpose, we have kept the values in MGCDB84 and Minnesota to their original values, as presented —and extensively used—by the authors of such databases. As last example, we want to discuss the DC13 subset in GMTKN55, which includes 13 problematic reactions for DFT methods. In the previous versions of the GMTKN database, the DC9 subset of Truhlar and co-workers was used. In GMTKN55, however, Goerigk and co-workers replaced it with a newer version that shares only one reaction with DC9, but it includes one reaction in common with the Styrene45 subset, and one in common with the C20C24 subset, both in MGCDB84.

For more details of all the subsets of every database, as well as references to their original sources, and a better overview of their overlaps, see also the Supporting Information.[7,12,22-27,30-40,46,48,51-174]

## Automation Software

ACCDB contains 10,049 geometry files—in *xyz* format, and appropriately named, including charge and spin multiplicity data—collected in one directory called "Geometries". Each file requires a single-point energy calculation, usually performed with quantum chemistry software engines, such as Gaussian[175] or Q-Chem.[176] These calculations will result in 44,931 unique reference data points, the majority of which are reaction energies (*vide supra*). The reference energies for each database or subset are reported in *csv* files that are available in either $E_h$, kcal $mol^{-1}$, or kJ $mol^{-1}$. Each reference file also includes the stoichiometry coefficients for the reaction in consideration, and reference to the corresponding filename in the "Geometries" directory.

As part of ACCDB, we provide a set of tools based on *snakemake* workflows,[177] that can be used for the automation of the jobs, the parsing of the output and reference files, and the collection of the final statistics. The automation files include: one *Snakefile*, with the *python* source code of the workflow; one configuration file (*config.yaml*), with user-specific configurations that can be simply specified using *yaml* syntax; and one template file, specific for each quantum chemistry software engine (sample *ginput.tmpl* and *qhcem.tmpl* files are provided for Gaussian and Q-Chem, respectively, extension to other programs are straightforward). The lists of the molecules pertaining to each database or set are also provided, and they are used in our workflow to extract the relevant *xyz* files from the "Geometries" directory. A representation of our software workflow is given in Figure 2. Such workflow will run all calculations on the selected databases (with the desired quantum chemistry engine), parse the output files of all completed calculations, and collect the results into a single *csv* file that can be used to calculate the statistics. More sophisticated statistical data can be collected from the output files, with simple modifications of the workflow. Instructions on how to interact with—and modify—the workflow, as well as details of all the configurations available in the *yaml* file are also included within the project. All relevant files are released under the GNU GPL license on Github.[178]

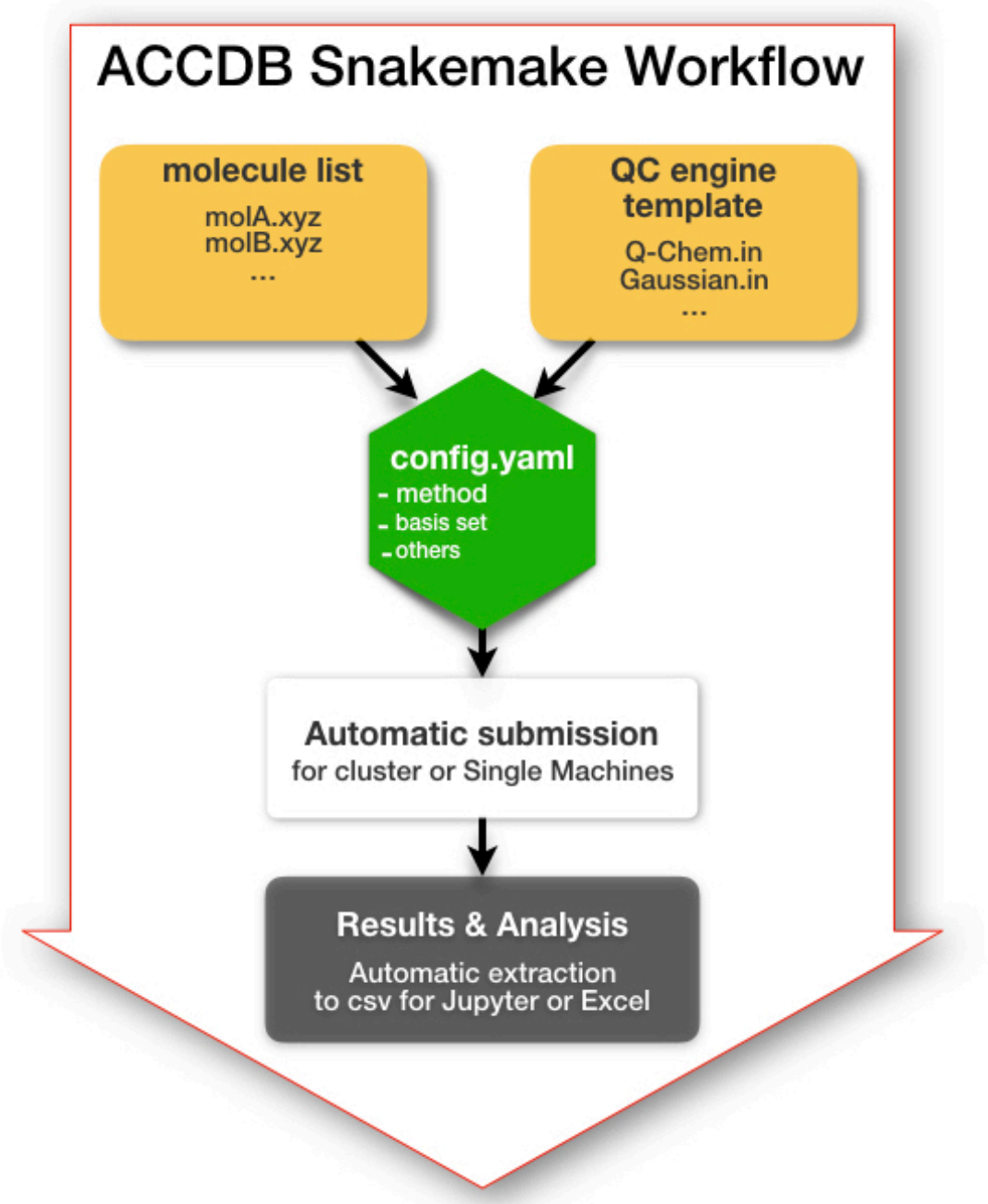

Figure 2. Pictorial representation of the ACCDB workflow.

We also want to point out that other automated procedures exist, but they significantly differ from ACCDB and the workflow presented above: de la Roza[179] gives the molecular geometries, reference data, and a pre-compiled Gaussian input file for every structure in the GMTKN55 database plus others, but his software lacks workflow controls and analysis tools; on the other hand, GC3PIE is a useful tool to submit jobs to cluster environments, [180] but it lacks an accompanying database of geometry files, and there is no trace of automatization. ACCDB is a much more general and flexible solution because it provides 10,049 geometry files (in a generic format that can be easily and automatically converted to any quantum chemistry engine input files via a simple customizable template file), a flexible workflow to automate the submission of the jobs and to control their execution from beginning to end (with tools applicable to both cluster environments, or single machines), as well as a parser for the retrieval of the final results, and for the calculation of the statistics on almost 45k reference data points (all of them also provided).

## Case Study: ACCDB on AMD Zen-based CPUs

As a simple case study, we used the entire ACCDB for Hartree–Fock (HF) calculations, in conjunction with the simple 3-21G basis set, as a benchmark for testing the performance of a set of commercially available processors of the Zen family, recently introduced onto the market by AMD. We chose HF because it is the most routine self-consistent-field method in computational chemistry, while we used the 3-21G minimal basis set because we needed a small basis set (in order to keep the overall time of completion of each calculation manageable), and because it is defined for most of the periodic table. The calculations range in size from H atom (3 basis functions), to a complex of Ru in MOR41 with 120 atoms (640 basis functions). It's important to remark that the purpose of this test is not to evaluate the accuracy of the method/basis-set combinations, but rather the speed of the CPUs. Hence, we will not report the accuracies of the calculations with respect to the unique reference energies.

We used three newly acquired (August 2018) machines equipped with similar hardware, but three different Zen-family CPUs. The processors have all been recently released (between 2017 and 2018), they have similar price per thread, and they cover different categories in the high-end market range of AMD Zen-based CPUs: the consumer category (Ryzen 7 2700X), the prosumer category (Ryzen Threadripper 1950X), and the professional category (Epyc 7281). The configuration for each machine is as similar as possible, and all machines are mounted on the same rack: each machine has 1 GB of RAM per thread, comparable motherboards (except for the Epyc processor, which is in a 2xCPU motherboard), the same operating system (Ubuntu server 18.04), and the same quantum chemistry engine (Q-Chem 5.1), with all calculations running on one single thread. On the one hand, launching multiple single-thread calculations in parallel (*vs.* single multiple-thread calculations in sequence) gives us the chance to measure the true scaling performance of each processor (*vs.* the scaling performance of the quantum chemistry engine). On the other hand, averaging out on almost 10,000 structures eliminates the risk of having a single slow calculation that bottlenecks the overall time. Comparable benchmarks could have been achieved using a single—sufficiently large—calculation ran in parallel multiple times, but using the entire database gives us the opportunity to understand the behavior of the CPUs under conditions that are closer to a real day-to-day research environment, and to extrapolate "educated guesses" on the timings for more advanced methods and basis sets combinations, by just adding the scaling information of the quantum chemistry engine.

***Single-thread Speed.*** We analyze first the single-thread performance of each CPU, by collecting the time (in minutes) it takes to run the entire ACCDB database on each machine with individual calculations running in sequence on a single-thread (ACCDB-st in Table 2). This number shows that the faster CPU for single-thread performance is the Ryzen 7, with the Epyc coming last. Perhaps not surprisingly, these numbers correlate well with the base clock rates of each CPU. Unfortunately, though, ACCDB-st doesn't represent parallel performance (multi-thread scaling), which is more important for evaluating the performance of CPUs in research environments, since the vast majority of calculations are usually performed in parallel on multiple threads/cores.

Table 2. Summary of the different machines equipped with AMD CPUs, and single-thread benchmark results.

| Machine | CPU | Cores/ Threads | Base Clock Rate (GHz) | ACCDB-st (min) |
|---|---|---|---|---|
| Ryzen | Ryzen 7 2700X | 8/16 | 3.7 | 974 |
| Threadripper | Ryzen Threadripper 1950X | 16/32 | 3.4 | 1200 |
| Epyc | 2xEpyc 7281 | 32/64[a] | 2.1 | 1508 |

[a] 16/32 per CPU in a dual-CPU server configuration.

***Multi-thread Scaling.*** In order to seek for the best performer with optimal compromise between single-thread speed and multi-thread scaling, we expanded our calculations on the entire ACCDB to include parallel calculations on each CPU. We started from ACCDB-st, with single calculations ran in sequence (1x), and performed subsequent runs of the full database doubling the number of calculations ran in parallel at each run (2x -> 4x -> 8x -> *etc.*), until full load is reached for each machine (again, each individual calculation is always a simple single-thread Q-Chem calculation, what changes is the number of simultaneous Q-Chem calculations on different threads). These detailed scaling results are reported in Table 3, as well as in the plots of Figure 3. Our best overall result for the entire ACCDB test is obtained with Threadripper at 16x (174 minutes), however its relative improvement over the single-thread performance—which ideally should be close to 16 for this case—is only 6.9. We found degradation of performance when higher number of threads are used with every processor. The relative improvement (R.I.) starts to deviate for its ideal value surprisingly early for both Ryzen, and Threadripper (the optimal ratio for both is only up to 4x).

Table 3. Multi-thread scaling performance of each CPU.[a]

| | Ryzen | | Threadripper | | Epyc | |
|---|---|---|---|---|---|---|
| Load:[b] | Time [min] | R.I.[c] | Time [min] | R.I.[c] | Time [min] | R.I.[c] |
| 1x | 974 | - | 1200 | - | 1508 | - |
| 2x | 513 | 1.9 | 719 | 1.7 | 716 | 2.1 |
| 4x | 262 | 3.7 | 306 | 3.9 | 346 | 4.4 |
| 8x | 235 | 4.1 | 256 | 4.7 | 233 | 6.5 |
| 16x | 236 | 4.1 | 174 | 6.9 | 238 | 6.3 |
| 32x | | | 186 | 6.4 | 237 | 6.4 |
| 64x | | | | | 247 | 6.1 |

[a] Time in minutes to run the entire ACCDB. [b] Full loads for each machine are: Ryzen 16x, Threadripper 32x, 2xEpyc 64x. [c] Relative improvement over single-thread performance.

The server processor Epyc does scale better up to 8x, but its poor single-thread performance is limiting its results significantly. The deviation of the R.I. from the ideal value at moderately high number of threads is an indication that the load that the quantum chemistry engine puts on the cores is very high, and the unexpected degradation of the performance at full-load puts into question the use of virtual cores for quantum chemistry calculations (results at 16x for Ryzen, at 32x for Threadripper, and at 64x for Epyc are all worse than the previous step for each processor).

***Is core virtualization useful for quantum chemistry?*** Simultaneous multi-threading (SMT, sometimes also called hyper-threading) is a popular way to increase the total number of cores seen by the operating system, by virtualizing multiple threads on one physical core. Dual-threading (virtualization of two threads on a single physical core) is now becoming the *de facto* standard for all new commercial processors introduced onto the market by both Intel and AMD, but considering our multi-thread scaling results reported above, a reasonable question arises: should we use SMT for quantum chemistry calculations? In Figure 3 we report our results for multi-core scaling with virtualization turned off, compared with the previous results with virtualization on.

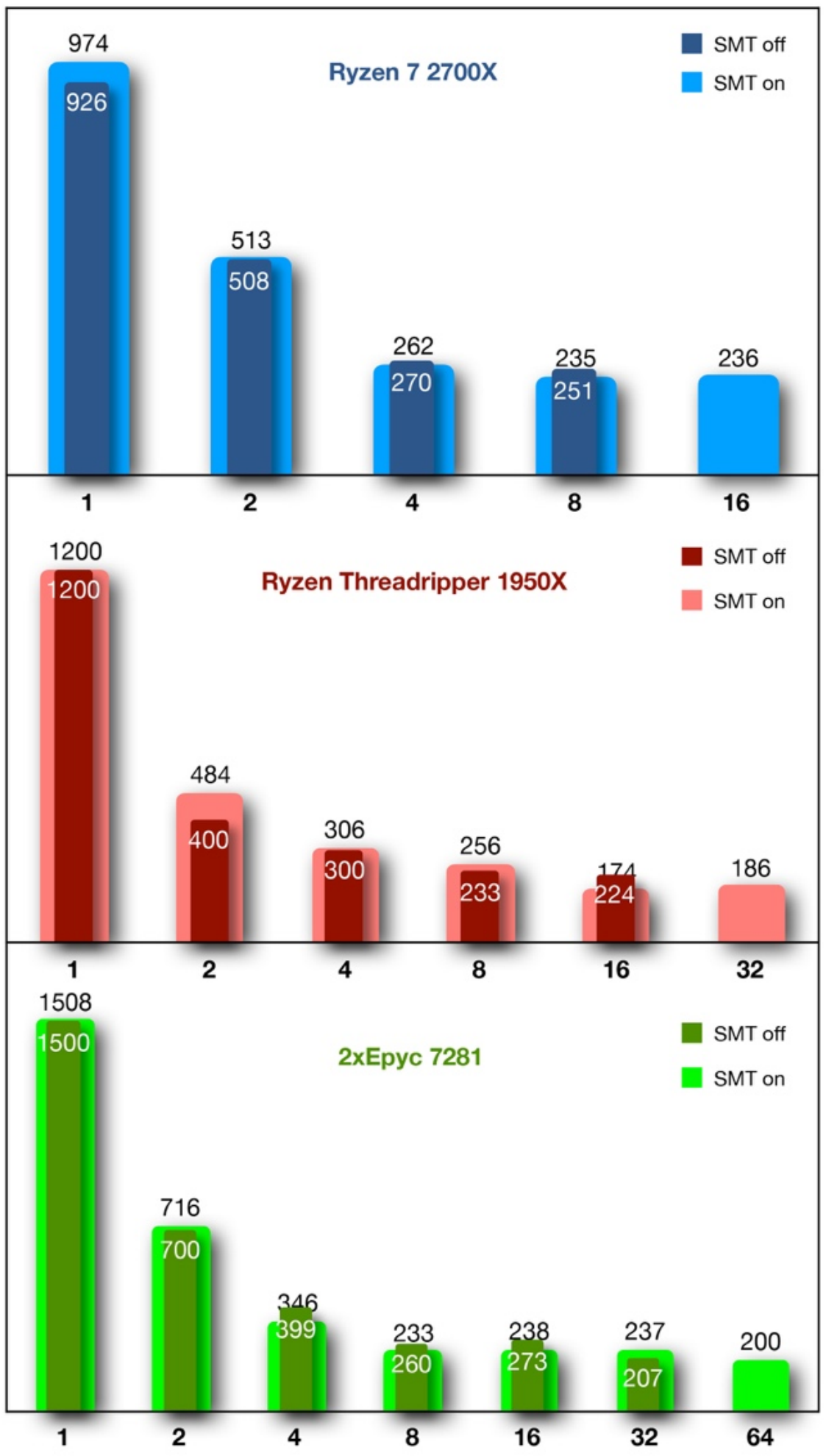

Figure 3. Parallel performance of the three CPUs under investigation (Ryzen 7: Blue, top panel; Threadripper: Red, middle panel; Epyc: Green, bottom panel), with simultaneous multi-threading (SMT) turned on (lighter bars, background) or off (darker bars, forefront), as a function of the number of threads/cores (the value of our ACCDB multi-thread scaling benchmark is also reported on each bar).

Despite the obvious loss of half of the threads of each CPU, it seems clear that the results don't change much, as long as at least half of the virtual threads are empty. The main difference is at full load, for which the processors with SMT turned on have a sensible degradation of the results. In light of these results, the answer to the question that we pose as the title of this section is rather simple: SMT does not present any advantage for quantum chemistry calculations, and we suggest to turn it off. (This is just the simplest strategy to avoid overfilling of threads, with the resulting significant degradation of the performance.) For machines where SMT is turned on by default—for example on shared computers or supercomputer centers—the best scaling performance can be achieved when at least half of the virtual threads for each physical core are left empty, and particular care should be given to not allow those virtual threads to perform any work (e.g. by reserving twice the amount of threads than those effectively used by the calculations).

***CPUs Recommendations.*** For calculations that are generally small (i.e. small molecules, small basis sets), the less expensive Ryzen 7 is the best choice among the CPUs we tested, mostly because its single-thread performance is the best, and because it scales reasonably well up to 4 threads/cores. Interestingly enough, because of the higher single-thread speed, the performance of Ryzen 7 on our benchmark at 4 threads, is similar to that of Threadripper at 8 threads. Threadripper becomes a feasible choice only for overall computation time, since our best result on the ACCDB benchmark was obtained with Threadripper at 16 threads (174 minutes). However, the relative improvement over the single-thread performance is not quite as good as we expected, and if 16 cores are not needed on one machine by the quantum chemistry engine itself, spreading the calculations on four independent Ryzen 7 machines working at four threads each, will result in less than half overall computation time over the best Threadripper result (with a moderately higher organizational effort to split the calculations, but at much lower retail prices). If calculations are large enough to require multiple threads on a single machine, the Epyc CPUs become competitive, because they allow to build individual machines with a significantly higher number of cores, and they have slightly better scaling performances than the two consumer CPUs we tested. Finally, in regards with virtualization of threads, we suggest to turn SMT off for quantum chemistry calculations, in machines where this can be done. In cases where this is not possible—for example on a shared cluster—the strategy that will provide the best results is to request twice the amount of virtual threads, and leave half of them unused.

## Conclusions

In the present article we introduced a large collection of computational chemistry databases (ACCDB), and the software tools that can be used to interact with them. ACCDB includes 44,931 unique reference data points, all at a level of theory significantly higher than DFT. The data covers all first four rows of the periodic table (H through Kr), most of the fifth row (W, Re, Ir, Pt, Au, Pb, Bi), and some actinides (Th through Cm). ACCDB is composed of five databases taken from literature: GMTKN, MGCDB84, Minnesota, DP284, and W4-17, plus two newly developed reaction energy databases, presented here for the first time and called W4-17-RE and MN-RE, and a collection of databases containing transition metals, also new to this article.

A set of expandable software tools for the interaction with ACCDB, its manipulation, and calculation of statistical data, is also presented here for the first time, and based on the *snakemake* workflow language.

Our case study also provides important insights on the performance of modern AMD CPUs, for routine quantum chemical calculations. The main results can be summarized as follows: when single-thread performance, or moderate scaling ability is required (small calculations), Ryzen 7 CPUs are the best choice. For very large calculations, where a high number of cores is required on an individual machine, the Epyc processors have a clear advantage. Despite Threadripper being the overall fastest processor in our benchmarks, it is hard to recommend it for quantum chemistry calculations, mostly because better results can be achieved with multiple (less-expensive) individual Ryzen 7 machines, or with a more tailored usage of one Epyc server. Despite the apparent advantage of doubling the number of available threads, the use of simultaneous multi-threading (virtualization of cores) is highly discouraged on all tested CPUs.

Finally, ACCDB is made available to the community, in the hope that it will be useful for different applications in many areas of computational chemistry, including development of new semi-empirical methods, and assessment of existing ones.

## Acknowledgments

The authors are grateful to the many scientists who made our aggregation of such a large number of high-level calculated data possible, with special shout-outs to Donald G. Truhlar, Martin Head-Gordon, Stefan Grimme, Jan M.L. Martin, Amir Karton, Lars Goerigk, and their entire groups. R.P. is also grateful to Narbe Mardirossian, Haoyu Yu, Xiao He, and Pragya Verma for their help in locating the correct structures for some subsets, and to Vladimir Konjkov for the suggestion of *snakemake* as a suitable tool for the development of the workflow.

**Keywords:** Database, Benchmarks, DFT, WFT, semi-empirical methods

Additional Supporting Information may be found in the online version of this article

Supporting Information for

# "ACCDB: A Collection of Chemistry DataBases for Broad Computational Purposes"

P. Morgante, R. Peverati
Chemistry Program, Florida Institute of Technology
150 W. University Blvd., 32901 Melbourne FL, United States.

Email: rpeverati@fit.edu

**Table S1. Databases included in ACCDB with their respective subsets and relevant citations.**

| GMTKN55 | | MGCDB84 | | Minnesota | Metals&EE |
|---|---|---|---|---|---|
| W4-11(TAE140)[1] | HAL59[49,50] | A24[62] | H2O16Rel5[94] | SRM2[106] | 3d-SSIP30[118] |
| G21EA[2] | AHB21[51] | DS14[63] | H2O20Rel10[80] | SRMGD5[106,107,134] | 4d-SSIP24[120] |
| G21IP[2] | CHB6[51] | HB15[64] | H2O20Rel4[8,46,85,86] | 3dSRBE2[108] | AIP28[137] |
| DIPCS10[3] | IL16[51] | HSG[44,65] | Melatonin52[58] | SR-MGN-BE107[106,109,110] | LTMBH26[138] |
| PA26[3,4,5] | IDISP[8,9,33,52,53] | NBC10[44,66-68] | YMPJ519[55] | ABDE13[111] | MOR41[139] |
| SIE4x4[3] | ICONF[3] | S22[43,44] | EIE22[95] | MR-MGM-BE4[107] | p-VR17[121] |
| ALKBDE10[6] | ACONF[54] | X40[50] | Styrene45[24] | MR-MGN-BE17[106] | Por21[140,141,148] |
| YBDE10[3,7] | AMINO20x4[55] | A21x12[69] | DIE60[32] | 3dSRBE4[108] | TMBH23[142-144] |
| AL2x6[3] | PCONF21[56,57] | BzDC215[70] | ISOMERIZATION20[1] | SRMBE10[106,134] | |
| HEAVYSB11[3] | MCONF[58] | HW30[71] | C20C24[26] | PdBE2[112] | **W4-17** |
| NBPRC[8,9,10] | SCONF[8,59] | NC15[72] | AlkAtom19[91] | FeCl[113] | TAE203[145] |
| ALK8[3] | UPU23[60] | S66[45,73] | BDE99nonMR[1] | 3dMRBE6[108] | BDE99[1] |
| RC21[3] | BUT14DIOL[61] | S66x8[45] | G21EA[2,8] | MRBE3[106] | HAT707[1] |
| G2RC[3,11] | | 3B-69-DIM[74] | G21IP[2,8] | CuH, VO, CuCl, NiCl[113] | ISOMERIZATION20[1] |
| BH76RC[3,8,12,13] | **MB08-165** | AlkBind12[75] | TAE140nonMR[1] | MR-TMD-BE3[106,114] | SN13[1] |
| FH51[14,15] | MB08-165[27] | CO2Nitrogen16[76] | AlkIsod14[91] | BH76[12,13,106,135] | |
| TAUT15[3] | | HB49[77-79] | BH76RC[8,12,13] | NCCE23[44,106,115-117,136] | **DP284** |
| DC13[3,8,16-26] | | Ionic43[51] | EA13[96] | CT7[106] | Dip152[146] |
| MB16-43[3,27] | | H2O6Bind8[80,81] | HAT707nonMR[1] | S6x6[45] | Pol132[147] |
| DARC[3,8,28] | | HW6Cl[80,81] | IP13[96] | NGDWI21[105,106] | |
| RSE43[3,29] | | HW6F[80,81] | NBPRC[8,9,10] | 3dEE8[114,118,119] | **New RE databases:** |
| BSR36[3,30,31] | | FmH2O10[80,81] | SN13[1] | 4dAEE5[120] | W4-17-RE[148] |
| CDIE20[32] | | Shields38[82] | BSR36[3,8,30,31] | pEE5[121] | MN-RE[148] |
| ISO34[33] | | SW49Bind345[83] | HNBrBDE18[97] | 4pISOE4[122] | |
| ISOL24[9,34] | | SW49Bind6[83] | WCPT6[41] | 2pISOE4[122] | |
| C60ISO[35] | | WATER27[8,46] | BDE99MR[1] | ISOL6/11[106,123] | |
| PARel[3] | | 3B-69-TRIM[74] | HAT707MR[1] | πTC13[5,106,124,125] | |
| BH76[3,12,13] | | CE20[40,84] | TAE140MR[1] | HC7/11[106,126] | |
| BHPERI[8,36-38] | | H2O20Bind10[80] | PlatonicHD6[98] | EA13/03[96,106,109,124,127,128] | |
| BHDIV10[3] | | H2O20Bind4[8,46,85,86] | PlatonicID6[98] | PA8[5,106] | |
| INV24[39] | | TA13[87] | PlatonicIG6[98] | IP23[96,106,109,124,127-129] | |
| BHROT27[3] | | XB18[49] | PlatonicTAE6[98] | AE17[16,104,106] | |
| PX13[40] | | Bauza30[88,89] | BHPERI26[8,99] | SMAE3[130,131,132] | |
| WCPT18[41] | | CT20[90] | CRBH20[100] | DC9/12[106,133] | |
| RG18[3] | | XB51[49] | DBH24[101,102] | | |
| ADIM6[3,42] | | AlkIsomer11[91] | CR20[103] | | |
| S22[43,44] | | Butanediol65[61] | HTBH38[13] | | |
| S66[45] | | ACONF[8,54] | NHTBH38[12] | | |
| HEAVY28[3,42] | | CYCONF[8,92] | PX13[40,84] | | |
| WATER27[46,47] | | Pentane14[93] | WCPT27[41] | | |
| CARBHB12[3] | | SW49Rel345[83] | AE18[104] | | |
| PNICO23[3,48] | | SW49Rel6[83] | RG10[105] | | |

**Table S2. Super position of subsets with at least partial overlap across all different databases.**
(Yellow indicates partial overlap, orange indicates complete overlap)

| GMTKN55[a,b] | MGCDB84[a] | W4-17 | Minnesota | Metals&EE |
|---|---|---|---|---|
| W4-11 (TAE140) | TAE140 | TAE203[c] | | |
| | BDE99, HAT707, ISOMERIZATION20, SN13 | BDE99, HAT707, ISOMERIZATION20, SN13 | | |
| G21EA | G21EA | | | |
| G21IP | G21IP | | | |
| PA26 | | | PA8 | |
| NBPRC | NBPRC | | | |
| BH76RC | BH76RC | | | |
| DC13[d] (DC9 in GMTKN30) | C20C24, Styrene45[d] | | DC9[d] | |
| MB16-43[e] (similar to MB08-165) | | | | |
| BSR36 | BSR36 | | | |
| CDIE20 | DIE60 | | | |
| ISOL24 (ISOL22 in GMTKN30) | | | IsoL6/11 | |
| BH76 | HTBH38, NHTBH38 | | BH76 | |
| BHPERI26 | BHPERI | | | |
| PX13 | CE20, PX13 | | | |
| WCPT18 | WCPT6, WCPT27 | | | |
| S22 | S22, HSG, NBC10 | | NCCE23 | |
| S66 | S66, S66x8 | | S6x6 | |
| WATER27 | WATER27, H2O20Bind4, H2O20Rel4 | | | |
| HAL59 | X40, XB51, XB18 | | | |
| AHB21, CHB6, IL16 | Ionic43 | | | |
| ACONF | ACONF | | | |
| Amino20x4 | YMPJ519 | | | |
| CYCONF (GMTKN30) | CYCONF (from GMTKN30) | | | |
| MCONF | Melatonin52 | | | |
| But14diol | Butanediol165 | | | |
| | RG10 | | NGDWI21 | |
| | IP13 | | IP23 | |
| | EA13 | | EA13 | |
| | AE18 | | AE17 | |
| | DBH24 | W4-08[f] | | |
| | | | 3dEE8 | 3d-SSIP30 |
| | | | 4dAEE5 | 4d-SSIP24 |
| | | | pEE5 | p-VR17 |

[a] MGCDB84 and GMTKN55 share a considerable number of subsets since some of the subsets in MGCDB84 are taken from GMTKN30. [b] For GMTKN55, all reference energies have been recalculated, and they differ (in general) from the ones reported in MGCDB84, or in Minnesota. [c] TAE203 replaces TAE140. The latter is taken from W4-11, together with BDE99, HAT707, ISOMERIZATION20, and SN13. [d] One of the reactions included in DC13 comes from C20C24, one from Styrene45, one from DC9, the others from literature. [e] MB16-43 substitutes MB08-165. [f] The reference energies for DBH24 are taken from W4-08 (ref. 102), but they are not included in our version of W4-17.

# References for all the subsets included in ACCDB.